\documentclass[]{siamart1116}
\usepackage{hyperref}

\usepackage{booktabs,cite}
\usepackage{mathtools}
\usepackage{subfig}

\usepackage{amsmath}
\usepackage{amssymb}

\numberwithin{theorem}{section}

\usepackage[american]{babel}
\usepackage[utf8]{inputenc}
\usepackage[T1]{fontenc}

\newcommand{\R}{\mathbb{R}}

\newcommand{\e}{\varepsilon}

\newcommand{\er}{\tilde\varepsilon_{r}\,}
\newcommand{\mur}{\mu_{r}\,}
\newcommand{\muri}{\mu_{r}^{-1}\,}
\newcommand{\ssr}{\sigma^\Sigma_r}
\newcommand{\tissr}{{\breve\sigma}^\Sigma_r}
\newcommand{\km}{k_{m}}
\newcommand{\kmr}{k_{m,r}}

\renewcommand{\vec}[1]{\boldsymbol{#1}}

\newcommand{\vB}{\vec B}
\newcommand{\vE}{\vec E}

\newcommand{\vME}{\vec{\mathcal{E}}}
\newcommand{\vMB}{\vec{\mathcal{B}}}

\newcommand{\vn}{\vec \nu}
\newcommand{\vp}{\vec \varphi}
\newcommand{\vx}{\vec x}
\newcommand{\vz}{\vec z}

\newcommand{\hQ}{\widehat{Q}}

\newcommand{\dx}{\,{\mathrm d}x}
\newcommand{\dox}{\,{\mathrm d}o_x}

\newcommand{\TheTitle}{Generation of surface plasmon-polaritons by edge effects}
\newcommand{\TheAuthors}{M. Maier, D. Margetis, M. Luskin}

\headers{\TheTitle}{\TheAuthors}

\title{{\TheTitle}\thanks{Submitted to the editors DATE.
\funding{The first and third authors (MM and ML) were supported in part by
  ARO MURI Award W911NF-14-1-0247. The second author’s research was
  supported in part by NSF DMS-1412769.}}}

\author{
  Matthias Maier\thanks{School of Mathematics, University of Minnesota
    Twin-Cities, Minneapolis, Minnesota 55455, USA
    (\email{msmaier@umn.edu},
    \email{luskin@umn.edu}).}
  \and
  Dionisios Margetis\thanks{Department of Mathematics, and Institute for
    Physical Science and Technology, and Center for Scientific Computation
    and Mathematical Modeling, University of Maryland, College Park,
    Maryland 20742, USA (\email{dio@math.umd.edu})}
  \and
  Mitchell Luskin\footnotemark[1]
}

\begin{document}


  \maketitle

  \begin{abstract}
    By using numerical and analytical methods, we describe the generation
    of fine-scale lateral electromagnetic waves, called {\em surface
    plasmon-polaritons} (SPPs), on atomically thick, metamaterial
    conducting sheets in two spatial dimensions (2D). Our computations
    capture the two-scale character of the total field and reveal how each
    edge of the sheet acts as a source of an SPP that may dominate the
    diffracted field. We use the {\em finite element method} to numerically
    implement a variational formulation for a weak discontinuity of the
    tangential magnetic field across a hypersurface.
    An adaptive, local mesh refinement strategy based on a posteriori error
    estimators is applied to resolve the pronounced two-scale character of
    wave propagation and radiation over the metamaterial sheet. We
    demonstrate by numerical examples how a {\em singular geometry}, e.g.,
    sheets with sharp edges, and sharp spatial changes in the associated
    surface conductivity may significantly influence surface plasmons in
    nanophotonics.
  \end{abstract}

  \begin{keyword}
    Time-harmonic Maxwell's equations,
    finite element method,
    surface plasmon-polariton,
    weak discontinuity on hypersurface
  \end{keyword}

  \begin{AMS}
    65N30, 78M10, 78M30, 78A45
  \end{AMS}

\section{Introduction}
\label{sec:Intro}

Recently, advances in nanophotonics have been made possible through the
design of atomically thick materials, e.g., graphene, with tunable, novel
electronic structure and macroscopic
conductivity~\cite{castroneto09,geim13}. A far-reaching goal is
to {\em precisely control} coherent light at the nanoscale, in the
terahertz frequency range~\cite{zhang-book}. This objective can be pursued
by manipulation of the microscopic parameters of conducting sheets. Under
suitable conditions, the sheet may behave as a metamaterial, exhibiting a
dielectric permittivity with a negative real part as a function of
frequency. By appropriate current-carrying sources and geometry, one may
generate electromagnetic waves that propagate with relatively short
wavelength along the sheet~\cite{bludov13}. A celebrated type of wave,
linked to various technological applications, is the surface
plasmon-polariton (SPP)~\cite{bludov13,fordweber84,samaier07}.

The present paper focuses on the computation of time-harmonic SPPs
intimately related to geometric effects, specifically the presence of edges
on conducting films, in the frequency regime in which the metamaterial
character of the sheet is evident. Our goal is to demonstrate numerically how spatial
changes in the morphology and surface conductivity of the sheet may
amplify, or spoil, the observation of SPPs. To this end, we combine
computational tools that include: (i) an implementation of the {\em finite
element method} for the {\em discontinuity} of the tangential magnetic
field across the sheet which is inherent to a class of thin conducting
metamaterials; and (ii) an approximate solution by the {\em method of Wiener
and Hopf} for an integral equation describing the electric field tangent to
a semi-infinite sheet (``reference case'') \cite{MML-sapm}.

Our main results can be summarized as follows.

\begin{itemize}
  \item
    We formulate a two-dimensional (2D) model for fine-scale SPPs induced
    by edge discontinuities. A few prototypical geometries are investigated
    numerically. In particular, we describe SPPs generated by: the edge
    discontinuity of a semi-infinite sheet; the gap between two co-planar,
    semi-infinite sheets; and a resonant, finite conducting strip.
  \item
    For the numerics, we adapt a variational framework for the finite
    element treatment of wave propagation~\cite{MML-jcp} to the 2D setting of
    SPP generation by edge effects. The underlying weak formulation
    embodies a discontinuity of the tangential magnetic-field component
    across the sheet and point singularities at material edges or
    discontinuities of the surface conductivity.
  \item
    We validate our numerical treatment by comparison of numerics to an
    approximate solution for the reference case of the semi-infinite sheet.
    We verify the analytically predicted singular behavior of the electric
    field near the edge.
  \item
    Based on our numerical simulations, we predict means of enhancing the
    observation of SPPs. Specifically, we demonstrate that the presence of
    a highly conducting material in the gap between two co-planar,
    semi-infinite metamaterial sheets may lead to an increase of the SPP
    amplitude. We also show that in this setting the dependence of the SPP
    amplitude on the gap width is distinctly different from the case with
    an ``empty gap'', i.e., when the material of the gap is the ambient
    medium. In addition, we characterize resonances for SPPs on a finite
    strip numerically, and demonstrate that the SPP maximal magnitude as a
    function of the gap width is described by a Lorentzian function.
   \end{itemize}

It has been well known that, because of phase matching in electromagnetic
wave propagation, SPPs cannot be excited by plane waves incident upon an
(idealized) infinite conducting sheet~\cite{bludov13}. The
relatively large wavenumber of the desired SPP cannot be directly matched at
the interface. This limitation is usually remedied by introduction of suitably
tuned gratings or localized current-carrying sources such as dipole
antennas near resonance~\cite{bludov13,gonzalez14}.

In this paper, we use a finite element approach to explore an alternate
scenario: the generation of SPPs via diffraction of waves by material
defects such as sharp edges of conducting sheets. We point out that such
geometric singularities, edges, can induce SPPs under radiation by
appropriately polarized plane waves. Our numerical simulations involve
hypersurfaces embedded in a 2D space for the sake of computational ease.
However, our key conclusions can be generalized to higher dimensions. One
of these conclusions is that the induced SPP may dominate over the
diffracted field in a range of distances away from the edge.

Our work here forms an extension of the approach in~\cite{MML-jcp} to more
realistic, singular geometries. In~\cite{MML-jcp}, we introduced a
numerical framework for tackling propagation of SPPs on planar
hypersurfaces via the finite element method. Our formulation relies on a
variational statement that incorporates the weak discontinuity of the
tangential component of the magnetic field across the sheet. This approach
offers the advantage of local refinement based on {\em a-posteriori} error
control. The computational work reported in~\cite{MML-jcp} is restricted to
idealized geometries, aiming to introduce a general platform rather than
address realistic applications.

In the present paper, we take a major step closer to applications by
extending our finite element approach~\cite{MML-jcp} to settings with sharp
edges and gaps on metamaterial, conducting films. Our ultimate purpose is
to model and simulate the effect of defects on the generation and
propagation of SPPs. We validate our numerics via comparison to an
approximate analytical solution for the semi-infinite conducting sheet with
recourse to the Wiener-Hopf method of factorization~\cite{MML-sapm}. This
solution analytically reveals that the algebraically singular field near
the edge transcends to a slowly decaying, fine-scale SPP away from the
edge.

\subsection{Scope: Excitation of SPPs}
\label{subsec:excitation}

The excitation of SPPs on interfaces between metals and dielectrics has
been conceived as a means of confining and manipulating coherent light in
the infrared spectrum~\cite{fordweber84}. Surface plasmons form
the macroscopic manifestation of the resonant interaction of
electromagnetic radiation (photons) with electrons in the plasma of the
metal surface. SPPs, in particular, are loosely defined macroscopically as
waves that have a relatively short wavelength and decay slowly along the
interface. SPPs may offer significant technological advantages; for example,
improvement of the emission of light from corrugated metallic surfaces or
nanoparticles~\cite{gruhlke86,mertens07}.

The macroscopic theory of surface plasmonics usually relies on Maxwell's
equations, specifically notions of classical electromagnetic wave
propagation near boundaries. This subject has been the focus of systematic
investigations in the case with radiowaves propagating over the earth or
sea; see, e.g.,~\cite{king92,chew99}. A central concept is the surface or
lateral wave, which is confined closely to the boundary~\cite{king92}.
Typically, at radio frequencies the surface wave between air and earth has
a phase velocity approximately equal to the phase velocity in air.

In surface plasmonics in the terahertz frequency regime, however, the
constitutive relations and geometry of the associated materials become more
intricate. Hence, the character of the ensuing surface wave is
substantially different from that at radio frequencies~\cite{samaier07}.
Here, we view the SPP as a type of surface wave. Notably, the wavelength of
the SPP with transverse-magnetic polarization can plausibly be made much
smaller than the wavelength of a plane wave in free space at the same
frequency~\cite{bludov13,cheng13}.

An emerging question is the following. How can the geometry or surface
conductivity of a low-dimensional material be controlled to generate a
desirable SPP? A satisfactory answer to this question requires
understanding through reliable computations how geometry affects solutions
to a class of boundary value problems for Maxwell's equations. In these
problems, an atomically thick material, e.g., graphene, may introduce a
{\em jump} proportional to a local property such as conductivity in
components of the electromagnetic field tangent to the sheet. The coupling
of this type of boundary condition to singular geometries in 2D is the
subject of the present paper.

\subsection{Past computational approaches}
\label{subsec:past}

Computational methods for plasmonics appear to be tailored to specific
applications. Next, we provide a brief, non-exhaustive summary of the main
computational tools found in the existing literature.

First, we comment on analytical treatments. In relatively simple settings
with planar boundaries, dispersion relations of SPPs have been sought
analytically through the use of plane-wave excitations; see, e.g., the
recent review in~\cite{bludov13}. Interestingly, the corresponding
geometries lack singular regions, e.g., sharp edges. A notable exception
concerns a metallic contact modeled as a strip of fixed width on a
substrate in 2D~\cite{bludov13,satou07}. For this geometry, an analytical
solution has been found via solving an integral equation in the Fourier
domain for the electric field tangent to the strip in the simplified case
with an {\em electrically small} strip width~\cite{bludov13}. This solution
is not applicable to conducting sheets with edges, since in the latter case
the edge may not be treated as a perturbation. A numerical study of
electrically large metal contacts by spectral methods is
provided in~\cite{satou07}; however, the transition of the field from a
singular behavior near the edge to an SPP is not discussed in
this work~\cite{satou07}.

Other settings in plasmonics consist of dipole sources over infinite planar
boundaries; then, analytical and semi-analytical solutions are developed
via the Fourier transform of the field~\cite{hanson08}. In some exceptional
cases, when the dipole and observation point lie on the sheet, exact
evaluation of field components is possible~\cite{margetis15}. These
approaches have the merit of yielding features of SPPs inherent to the
nature of the point source; but they convey little or no information about
how realistic geometric effects related to the finite size of the sheet may
influence the SPP.

In more complicated geometries, the excitation of plasmons has been studied
under the ``quasistatic approximation'' in which the typical size of the
scatterer is much smaller than the wavelength of the incident radiation
field~\cite{samaier07}. However, this approximation is expected to be
questionable as the frequency becomes higher or the size of the scatterer
is comparable to the free-space wavelength~\cite{samaier07}.

A variety of numerical methods for nanophotonics have been reported in the
literature; for a recent review, see, e.g.,~\cite{gallinet15}. These
approaches include: the volume integral-equation
approach~\cite{kern11,kottmann00,kern09}, which exploits the integral form
of Maxwell's equations along with the respective Green's function in closed
form in the frequency domain; and the akin boundary element
method~\cite{garcia02,mayergoyz05,hohenester12}, which makes use of Dirac
masses as test functions and often employs the scalar and vector
potentials. Numerical methods of different nature are the finite-difference
time-domain method~\cite{mohammadi08,liu09} and the finite-difference
method~\cite{felsen-book}, which invoke space discretization and the
respective approximation of the electromagnetic field by piecewise-constant
functions. This type of approximation may be challenged in plasmonics,
where the electromagnetic field may vary appreciably over short distances.
An improvement has been offered by the finite element
method~\cite{monk03,webb99,chew99} with certain choices of basis functions,
for example, divergence-free functions. An issue of importance is to
satisfy the radiation condition far away from the source. This can be
accomplished in the numerical scheme via the notion of the perfectly
matched layer (PML)~\cite{berenger94,chew94}. In passing, we should also
mention the discrete dipole approximation~\cite{devoe64,dagostino13} by
which the (continuum) scatterer is replaced by a finite array of
polarizable particles or dipoles; this technique can be viewed as the
outcome of discretization of the volume integral equation.

\subsection{Our computational treatment}
\label{subsec:approach}

Recently, we formulated a variational framework for the finite element
treatment of wave propagation along metamaterial conducting sheets embedded
in spaces of arbitrary dimensions~\cite{MML-jcp}. Our formulation
incorporates a weak discontinuity of the tangential component of the
magnetic field; this jump is responsible for the fine scale of the SPP. The
corresponding discretization scheme was implemented with a modern finite
element toolkit \cite{dealii84}, and accounts both for the small wavelength
of the SPP as well as for the radiation condition at infinity via a PML.
This approach allows for error control and, thus, an efficient numerical
approximation of the underlying boundary value problem. In our past
work~\cite{MML-jcp}, we validated this approach by numerical simulations
restricted to settings with dipoles over infinite planar geometries in 2D.

With the present paper, our goal is to adapt the aforementioned approach to
more realistic settings and real-world applications. Hence, we extend our
finite element treatment to more complicated geometries, especially
conducting sheets with edges and gaps in 2D. We demonstrate via simulations
that our variational framework and numerical implementation by appropriate
curl-conforming Nédélec-elements correctly produces the edge singularity of
the electric field. Our numerics for finite sheets show how the diffracted
electric field transcends from a singular behavior near each edge to the
SPP away from the edge. We validate our numerical approach by use of a
semi-infinite sheet: in this case, our numerical results  are in excellent
agreement with the analytical prediction~\cite{MML-sapm}. This study places
on a firm foundation our finite element approach for SPPs generated and
sustained  by defects and finite-size effects.

\subsection{Pending issues}
\label{subsec:pending}

Our work here leaves a few open problems for near-future study. For
example, we do not solve the related boundary value problem for Maxwell's
equations in three spatial dimensions (3D). This important topic is the
subject of work in progress. Furthermore, in our model we use homogeneous
and isotropic material parameters. The case with spatially varying conductivity of the sheet, where the SPP may be the result of
homogenization~\cite{cheng13}, lies beyond our present scope. In a similar
vein, we have not made any attempt to study more complicated forms of
discontinuities for the electromagnetic-field components across the sheet,
say, in the presence of a magneto-electric effect~\cite{zulicke14}. We should also mention the
experimentally appealing case with a receiving antenna lying on
graphene~\cite{gonzalez14}.
  This problem can be described by our variational approach, and is the
  subject of future research as well.

Note that surface plasmonics comprise a class
of multiscale problems: the electronic structure of low-dimensional
materials should be linked to the phenomenology of SPPs. Here, we invoke a
macroscopic model of SPPs via Maxwell's equations. The emergence of the
related phenomenology from microscopic  principles needs to be
further understood.

\subsection{Paper organization}
\label{subsec:organiz}

The remainder of this paper is organized as follows. In
section~\ref{sec:modelandmethods}, we introduce the relevant boundary value
problem and summarize analytical results for the reference
case~\cite{MML-sapm}. Section~\ref{sec:numerics} focuses on our numerical
approach; in particular, the use of a PML and adaptive local refinement for
good resolution of the SPP is discussed. In Section~\ref{sec:computation},
we present and discuss computational results for the following geometries:
a semi-infinite sheet (reference case); two co-planar, semi-infinite sheets
with a gap; and a resonant, finite conducting strip.
Section~\ref{sec:conclusion} concludes our paper with a summary of the main
findings.


\section{Model and analytical results}
\label{sec:modelandmethods}

In this section, we introduce the main ingredients of the model and
analytical approach: a boundary value problem for Maxwell's equations,
which incorporates a discontinuity for the magnetic-field component tangent
to an arbitrary conducting sheet in 2D (section~\ref{subsec:bvp}); and an
analytical formula for the SPP on a semi-infinite sheet
(section~\ref{subsec:integral_eqn}). Our formula manifests the fine scale
of the SPP if the sheet conductivity satisfies a certain condition
consistent with the metamaterial nature of the sheet. In this vein, we
describe solutions for the electric field far from and near the edge on a
semi-infinite sheet \cite{MML-sapm} (section~\ref{subsec:integral_eqn}).

\subsection{Boundary value problem and geometry}
\label{subsec:bvp}

\begin{figure}[t]
  \centering



  \includegraphics{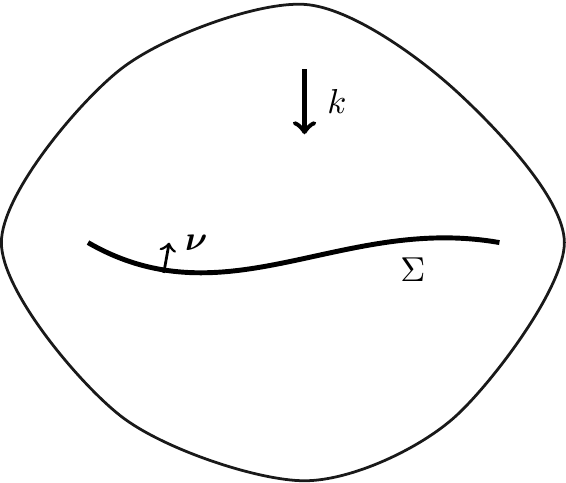}
  \caption{Schematic of the geometry. A plane wave with wave vector $k$ is
    incident upon an (arbitrarily oriented) interface $\Sigma$ of unit normal $\nu$ in
    an unbounded domain, in order to excite an SPP.}
  \label{fig:bvp}
\end{figure}

Following~\cite{MML-jcp}, we formulate a boundary value problem for
Maxwell's equations by including a conducting sheet in an unbounded space;
see Figure~\ref{fig:bvp}. The starting point is the strong form of
Maxwell's equations for the time-harmonic electromagnetic field
$(\vME(\vx,t), \vMB(\vx,t))$, viz.,
\begin{align*}
  (\vME(\vx,t), \vMB(\vx,t))=\text{Re}\,\big\{e^{-i\omega t}(\vE(\vx),
  \vB(\vx))\big\}~.
\end{align*}
By splitting the electric and magnetic fields into the primary (incident)
field and the scattered field, $\vE=\vE^{\text{in}}+\vE^{\text{sc}}$,
$\vB=\vB^{\text{in}}+\vB^{\text{sc}}$, Maxwell's equations (in a
source-free form) for the scattered field outside the sheet are written
as~\cite{chew99}
\begin{align}
  \begin{cases}
    \begin{aligned}
      -i\omega\vB^{\text{sc}}+\nabla\times\vE^{\text{sc}} \;&= \;0,
      \\[0.1em]
      i\omega\tilde\e\vE^{\text{sc}}+\nabla\times\big(\mu^{-1}\vB^{\text{sc}}\big) \;&= \;0,
      \\[0.1em]
      \nabla\cdot\vB^{\text{sc}} \;&=\; 0,
      \\[0.1em]
      \nabla\cdot\big(\tilde\e\vE^{\text{sc}}\big) \;&=\; 0.
    \end{aligned}
  \end{cases}
  \label{eq:timeharmonicmaxwell}
\end{align}
Here, Gauss' law, expressed by the last two equations, is redundant.
In~\eqref{eq:timeharmonicmaxwell}, the material parameters are time
independent. The second-rank tensors $\mu(\vx)$ and $\tilde\e(\vx)$ denote
the {\em effective} magnetic permeability and complex permittivity of the
unbounded medium; in particular, $\tilde\e(\vx)=\e(\vx)+i
\sigma(\vx)/\omega$, where $\e(\vx)$ and $\sigma(\vx)$ are the usual
permittivity and conductivity of the medium.\looseness=-1

Equations~\eqref{eq:timeharmonicmaxwell} are enforced in suitable unbounded
regions of the Euclidean space $\mathbb{R}^n$ ($n=2,\,3$), by exclusion of
the conducting sheet. Hence, system~\eqref{eq:timeharmonicmaxwell} must be
complemented with the appropriate boundary condition across the sheet which
is represented by an oriented hypersurface $\Sigma$, $\Sigma\subset
\mathbb{R}^n$, with unit normal $\vec \nu$ and surface conductivity
$\sigma^\Sigma(\vx)$~\cite{bludov13}. For a conducting sheet, the
electromagnetic field satisfies the conditions~\cite{bludov13}
\begin{align}
  \begin{cases}
    \begin{aligned}
      \{\vec \nu\times\big[\big(\mu^{-1}\vB\big)^+-\big(\mu^{-1}\vB\big)^-\big]\}
      \Big|_{\Sigma}
      \;&=\;
      \{\sigma^\Sigma(\vx)\,[(\vec \nu \times\vE)\times\vec \nu]\}\Big|_{\Sigma}~,
      \\[0.2em]
      \{\vec \nu\times\big(\vE^+-\vE^-\big)\}\Big|_{\Sigma}
      \;&=\; 0~,
    \end{aligned}
  \end{cases}
  \label{eq:jumpcondition}
\end{align}
where $\vE^{\pm},\,\vB^{\pm}$ is the restriction of the field to each side
($\pm$) of the sheet and $\sigma^{\Sigma}(\vx)$ is a second-rank tensor
with domain on $\Sigma$.

In the presence of compactly supported external sources, e.g., electric or
magnetic dipoles,~\eqref{eq:timeharmonicmaxwell}
and~\eqref{eq:jumpcondition} must be complemented with a radiation
condition at infinity. For an isotropic medium, this condition takes the
form of a Silver-M\"uller condition~\cite{chew99}. Specifically, if $c$ is
the speed of light at infinity, we require that
\begin{equation}
  \lim_{|\vx|\to\infty}\left\{\vB^{\text{sc}}\times \vx -
  c^{-1}|\vx|\vE^{\text{sc}}\right\}=0~,\quad
  \lim_{|\vx|\to\infty}\left\{\vE^{\text{sc}}\times \vx +
  c|\vx|\vB^{\text{sc}}\right\}=0\quad (\vx\notin \Sigma)~.
\end{equation}

Motivated by the setting with an infinite planar sheet~\cite{bludov13}, we
will pay particular attention to material properties and external sources
that allow for the generation of a short-wavelength SPP on the conducting
sheet. Specifically, for an isotropic and homogeneous sheet we expect that
a necessary condition for such an SPP is $\text{Im}\,\sigma^\Sigma >0$. In
addition, the sheet needs to be radiated by an appropriately
(transverse-magnetic) polarized wave. We conclude this subsection with a
practically appealing definition~\cite{cheng13}:

\begin{definition}\label{def:nonr}
  For scalar $\sigma^\Sigma$, the {\rm nonretarded frequency regime} is
  characterized by
  \begin{equation}\label{eq:nonr}
    {\left|\frac{\omega\mu\sigma^\Sigma}{k}\right|}\ll 1~,
  \end{equation}
  i.e., a surface resistivity ($1/\sigma^\Sigma$) that is much
  larger in magnitude than the intrinsic impedance of the ambient medium.
\end{definition}

By recourse to an analytical solution for the reference case
(section~\ref{subsec:integral_eqn}), we will see that~\eqref{eq:nonr} along
with the condition $\text{Im}\,\sigma^\Sigma>0$ imply that an SPP is present
on a semi-infinite sheet and has a wavenumber, $\km$, with $|\km|\gg k$.
This property in turn yields two distinct length scales of wave propagation
in this problem.

\subsection{Reference case: Explicit formulas}
\label{subsec:integral_eqn}

\begin{figure}[!t]
  \centering
  \subfloat[]{
    \includegraphics{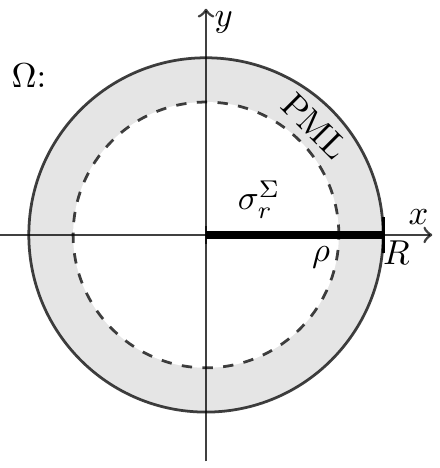}
  }
  \subfloat[]{
    \includegraphics{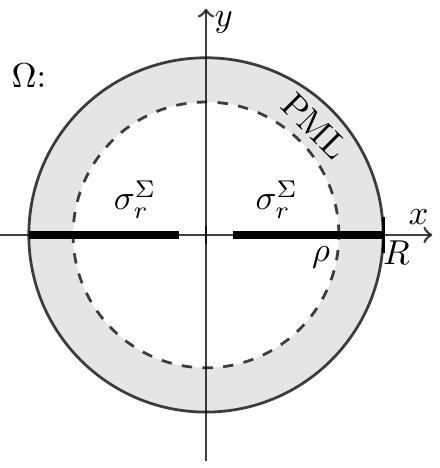}
  }

  \subfloat[]{
    \includegraphics{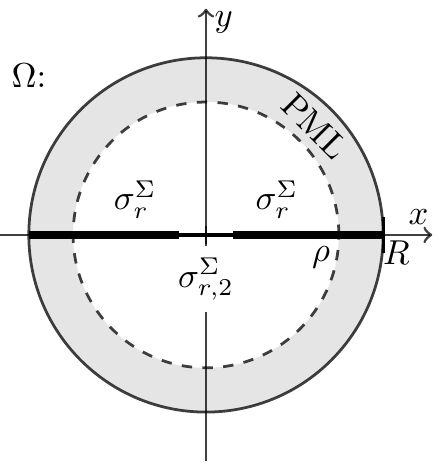}
  }
  \subfloat[]{
    \includegraphics{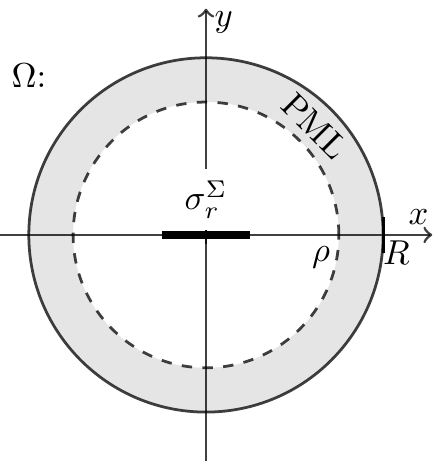}
  }
  \caption{
    Schematic of the computational domain, $\Omega$, of radius $R$ for
    different geometries with metamaterial sheet of conductivity
    $\sigma_r^\Sigma$ in our numerics. The related perfectly matched layer
    (PML), shown in grey, occupies a spherical shell of radii $\rho$ and
    $R$ ($\rho< R$). (a) Semi-infinite sheet configuration; (b) two
    semi-infinite, co-planar sheets, separated by an empty gap; (c) two
    semi-infinite, co-planar sheets separated by gap filled with a highly
    conducting material of conductivity $\sigma^\Sigma_{r,2}$,
    $\text{Re}\,\sigma^\Sigma_{r,2}\gg 1$; (d) finite strip.}
  \label{fig:geometries}
\end{figure}

In the case with a semi-infinite sheet, $\Sigma=\{(x,0)\in\R^2\,:\,x\ge0\}$
[see Figure~\ref{fig:geometries}(a)], Maxwell's equations
\eqref{eq:timeharmonicmaxwell} with jump condition \eqref{eq:jumpcondition}
admit a closed-form solution for the electric field, $E_x(x,0)$, tangential
to the sheet. This solution is obtained by application of the Wiener-Hopf
method to an integral equation~\cite{MML-sapm}. In the following we briefly
motivate and describe our formalism and  results; for a detailed discussion
we refer the reader to \cite{MML-sapm}.

The key ingredient of the analytical approach is the formulation of
\eqref{eq:timeharmonicmaxwell} and \eqref{eq:jumpcondition} in terms of an
integral equation for the (non-dimensional) $x$-component,
$u(x)=E_x(k^{-1}x,0)/E_0$, of the electric field on the sheet, where $E_0$
is a typical amplitude of the incident electric field. The governing
integral equation reads
\begin{equation}
  u(x) =
  u^{\rm in}(x) +
  \frac{i\omega\mu\sigma^\Sigma}{k}\left(\frac{d^2}{d x^2}+1\right)\int_0^\infty
  dx'\,\mathfrak K(x-x')\,u(x') \quad x>0~,
  \label{eq:n-integrod}
\end{equation}
where $u^{\rm in}$ corresponds to the incident field, $x$ is scaled by
$1/k$ where $k$ is real, and $\mathfrak K$ is an appropriate kernel, which
corresponds to a Green's function for the scalar Helmholtz equation; see
\cite[Sec.\,1]{MML-sapm}. By formally extending $u$ and $u^{\rm in}$ to the
whole real axis through setting $u(x)=u^{\rm in}(x)\equiv 0$ for $x<0$, and
introducing an unknown function, $g$, for consistency ($g(x)\equiv 0$ if
$x>0$), we write
\begin{equation}\label{eq:n-integrod-ext}
  u(x)-\frac{i\omega\mu\sigma^\Sigma}{k}\left(\frac{d^2}{d
  x^2}+1\right)\int_{-\infty}^\infty dx'\,\mathfrak K(x-x')\,u(x')=u^{\rm
  in}(x)+g(x)\quad x\in\mathbb{R}~.
\end{equation}
Equation~\eqref{eq:n-integrod-ext} is solved explicitly in terms of a
Fourier integral by use of the Wiener-Hopf method \cite{MML-sapm}. To this
end, we consider a plane wave as the incident field, $E_x^{\rm in}=E_0
e^{ix\,\sin\alpha}$, for some incidence angle, $\alpha$. In the following,
we summarize the main findings of this approach.

\paragraph{\rm\bf Field far from the edge}
In applications of plasmonics, the region consisting of points of the sheet
sufficiently far from the edge deserves some attention. This region can be
characterized by the condition $x\gg 1$. In this region, the SPP may
dominate over the diffracted field; the latter is defined as the scattered
field after removal of the direct reflection of the incident wave. The
solution to \eqref{eq:n-integrod-ext} can be written as
\begin{align}\label{eq:far-field-E}
  u(x) \;\approx\;&
  \left(1+\frac{\omega\mu\sigma^\Sigma}{2k}\cos\alpha\right)^{-1}\, e^{ix\sin\alpha}
  \\
  &-e^{-\hQ_+(\sin\alpha)+\hQ_+(\km/k)}e^{i(\km/k) x}~.
  \nonumber
\end{align}

A few remarks on~\eqref{eq:far-field-E} are in order. The first term
describes the sum of the incident field and its direct reflection from an
infinite sheet, while the second term is the SPP. The effect of the edge is
expressed by values of the {\em split function} $\hQ_+(\xi)$, which is
defined by the contour integral~\cite{MML-sapm}
\begin{equation}\label{eq:Q+-integral}
  \hQ_+(\xi)=\frac{\xi}{\pi
  i}\int_0^{\infty}\frac{d\zeta}{\zeta^2-\xi^2}\,
  \ln\Biggl(1+\frac{\omega\mu\sigma^\Sigma}{2k}\sqrt{1-\zeta^2}\Biggr)~,\qquad {\rm Im}\xi>0~,
\end{equation}
and can be evaluated approximately, in closed form, in the nonretarded
regime (Definition~\ref{def:nonr})~\cite{MML-sapm}. In principle, this
$\hQ_+(\xi)$ enters  the Fourier integral for the exact solution, $u$, for
$x>0$~\cite{MML-sapm}. In the region under consideration (if $x\gg 1$),
$\hQ(\xi)$ needs to be computed for particular values of $\xi$ in order to
compare formula~\eqref{eq:far-field-E} against numerical results of the
finite element method (section~\ref{sec:computation}). For example, for
$\alpha=0$, we compute $\hQ_+(\xi)$ for $\xi=0$ and $\xi=k_m/k$
approximately (see \cite[Sec. 3]{MML-sapm}), viz.,
\begin{align*}
  \hQ_+(0)\approx0.0005+0.05i,\qquad
  \hQ_+(k_m/k)\approx0.346574+0.392699i~.
\end{align*}

We should add that the SPP wavenumber, $\km$, obeys the dispersion relation
$k_{\perp}:=\sqrt{k^2-\km^2}=-2k^2/(\omega\mu\sigma^\Sigma)$, which
furnishes the wavenumber, $k_\perp$, of propagation transverse to the
sheet; thus, $\km/k=\sqrt{1-4k^2/(\omega\mu_0\sigma^\Sigma)^2}$, and
$|\km/k|\gg 1$ in the nonretarded regime
(Definition~\ref{def:nonr})~\cite{MML-sapm}. By imposing
$\text{Im}\,k_{\perp}>0$ according to the radiation condition at infinity
vertically to the sheet, we obtain $\text{Im}\, \sigma^\Sigma > 0$ which
expresses the metamaterial character of the sheet.

An additional contribution to $u(x)$, not included
in~\eqref{eq:far-field-E}, is the radiation field due to the edge. This
contribution is negligible in the region considered here and, thus, is
omitted.
\medskip

\paragraph{\rm\bf Near-edge field} Another region of significance is the
vicinity of the edge, where $|(\km/k) x|\ll 1$. In this region, the SPP
interferes with the incident, reflected and radiation fields to yield a
vanishing $x$-component of the (total) electric field on the sheet. Hence,
as $x\searrow 0$, the solution, $u$, of integral
equation~\eqref{eq:n-integrod-ext} cannot be separated into the distinct
(physical) contributions that are evident in the field far from the edge.
Instead, we analytically compute the asymptotic behavior~\cite{MML-sapm}
\begin{align}
  u(x) \;\approx\; \frac{2}{\sqrt{\pi}}\sqrt{\frac{2k}{\omega\mu\sigma^\Sigma}}\,e^{-i\pi/4}\,\sqrt{x}
  \qquad \mbox{as}\ x\searrow 0~.
  \label{eq:asymptotic0}
\end{align}
This formula manifests the singular behavior of the field at the edge.

%

\section{Numerical approach}
\label{sec:numerics}

In this section, we describe an implementation of the finite element method
in order to solve the boundary value problem of section~\ref{subsec:bvp}
based on curl-conforming Nédélec-elements~\cite{nedelec86}. First, we state
boundary-value problem \eqref{eq:timeharmonicmaxwell} in terms of a weak
formulation that embodies discontinuity \eqref{eq:jumpcondition} of the
tangential magnetic-field implicitly by means of an interior interface
integral. Second, we discuss a number of numerical aspects to solve the
weak formulation; in particular, the use of a specifically tuned PML, as
well as a local refinement strategy to resolve the SPP.

\subsection{Variational formulation and discretization}
\label{subsec:var-form}

By a standard manipulation in electromagnetic theory, the substitution of
the first equation of \eqref{eq:timeharmonicmaxwell} into the second one
yields a second-order partial differential equation for the vector-valued
electric field, $\vE$. To rescale the resulting equation to a desired,
dimensionless form, we choose to set $k=1$, or, equivalently, scale the
vector position $\vx$ by $1/k$, in the spirit of
section~\ref{subsec:integral_eqn}. Thus, we introduce the following related
scalings~\cite{MML-jcp}:
\begin{subequations}
  \label{eq:rescaling}
  \begin{gather}
    \vx \;\rightarrow\; k\,\vx,
    \quad
    \nabla \;\rightarrow\; \frac1{k}\,\nabla,
    \quad
    \kmr = \frac\km k,
    \\
    \mu \;\rightarrow\; \mur =\frac1{\mu_0}\mu,
    \quad
    \tilde\e \;\rightarrow\; \er = \frac1{\e_0}\tilde\e,
    \quad
    \sigma^\Sigma \;\rightarrow\; \ssr =
    \sqrt{\frac{\mu_0}{\e_0}}\,\sigma^\Sigma.
  \end{gather}
\end{subequations}
Accordingly, the equation for $\vE$ now reads
\begin{align}
  \nabla\times\big(\muri\nabla\times\vE\big)
  -\er\vE
  \;=\; 0.
  \label{eq:2ndorderequation}
\end{align}
The multiplication of the last equation with a test function $\vp$ and
subsequent integration by parts yield \cite{MML-jcp}
\begin{multline}
  \int_\Omega (\muri\nabla\times\vE^{\text{sc}})\cdot(\nabla\times\bar\vp)\dx
  \;\;-\;
  \int_\Omega(\er\vE^{\text{sc}})\cdot\bar\vp\dx
  \\
  -\,i\,\int_\Sigma(\ssr\vE^{\text{sc}}_T)\cdot\bar\vp_T\dox
  \;-\;
  i\,\int_{\partial\Omega}\sqrt{\muri\er}\vE^{\text{sc}}_T\cdot\bar\vp_T\dox
  \;=\;
  \\
  \,i\,\int_\Sigma(\ssr\vE^{\text{in}}_T)\cdot\bar\vp_T\dox~.
  \label{eq:variationalformulation2}
\end{multline}
where $E^{\text{sc}}$ is the scattered field and $E^{\text{in}}$ denotes
the incident field (section~\ref{subsec:bvp}). An appropriate trial and
test space for the weak formulation is
\cite{MML-jcp},\cite[Theorem\,4.1]{monk03}
  \begin{align*}
    \vec X(\Omega)=\Big\{\vp\in \vec L^2(\Omega)\;:\;
    \nabla\times\vp\in L^2(\Sigma)^3,\;
    \vp_T\big|_\Sigma\in L^2(\Sigma)^3,\;
    \vp_T\big|_{\partial\Omega}\in L^2(\partial\Omega)^3
    \Big\}~,
  \end{align*}
where $L^2$ denotes the space of measurable and square integrable
functions. By choosing this space, the formerly strong interface condition
\eqref{eq:jumpcondition} is now naturally embedded in the variational
formulation. In more detail, the statement
\begin{align*}
  \{\vec \nu\times\big(\vE^+-\vE^-\big)\}\Big|_{\Sigma} = 0
\end{align*}
is a consequence of $\vE\in\vec H(\text{curl};\Omega)$, and the jump
condition
\begin{align*}
  \{\vec \nu\times\big[\big(\mu^{-1}\vB\big)^+-\big(\mu^{-1}\vB\big)^-\big]\}
  \Big|_{\Sigma} \;&=\;
  \{\sigma^\Sigma(\vx)\,[(\vec \nu \times\vE)\times\vec \nu]\}\Big|_{\Sigma}
\end{align*}
is enforced by the term
$-\,i\,\int_\Sigma\ssr\vE_T\cdot\bar\vp_T\dox$
in the variational formulation \cite{MML-jcp}.

We implement \eqref{eq:variationalformulation2} by using curl-conforming
Nédélec-elements~\cite{nedelec86} with the help of the finite
element toolkit \textsc{deal.II}~\cite{dealii84}. The computational
domain $\Omega$ was discretized with a quadrilateral mesh. In order to
allow for local refinement, we use the well-known concept of hanging nodes
(see, e.\,g., \cite{carstensen09} for an overview) to relax the usual
mesh regularity assumptions~\cite{ciarlet02}. The resulting system of
linear equations is solved with the direct solver \textsc{Umfpack}
\cite{suitesparse4}.

\subsection{Perfectly matched layer for SPPs}
\label{subsec:pml}

Next, we discuss a construction of a PML~\cite{berenger94, chew94} for the
rescaled Maxwell equations with a jump condition in connection to the
boundary problem of section~\ref{subsec:bvp}~\cite{MML-jcp}. The concept of
a PML was pioneered by Bérenger~\cite{berenger94} and can be viewed as a
layer with modified material parameters ($\er$, $\mur$) placed near the
boundary of the computational domain; cf. Figure~\ref{fig:geometries}. The
core idea is to tune the material parameters inside the PML in such a way
that all outgoing electromagnetic waves decay exponentially with no
artificial reflection due to truncation of the domain. The PML is an
indispensable tool for truncating unbounded domains for time-harmonic
Maxwell's equations, and other, akin partial differential equations, and is
often used in the numerical approximation of scattering
problems~\cite{monk03,chew94,berenger94}.

We follow the approach to a PML for time-harmonic Maxwell's equations
discussed in~\cite{chew94}. The idea is to use a formal change of
coordinates from the computational domain $\Omega\subset\R^3$ with
real-valued coordinates to a domain
$\acute\Omega\subset\{z\in\mathbb{C}\::\:\text{Im}\,z\ge0\}^3$ with
complex-valued coordinates (and non-negative imaginary part)~\cite{monk03};
and then transform back to the real-valued domain. For simplicity, we
assume that the interface $\Sigma$ is parallel to the unit vector $\vec
e_r$ within the PML (i.e., the normal $\vn$ is orthogonal to $\vec e_r$).
For details, we refer the reader to \cite{MML-jcp}. This procedure results
in the following modified material parameters $\er$, $\mur$ and $\ssr$
within the PML:
\begin{align}
  \begin{cases}
    \begin{aligned}
      \muri
      &\quad\longrightarrow\quad
      \breve\mu_r^{-1} = B\muri A~,
      \\[0.1em]
      \er
      &\quad\longrightarrow\quad
      \breve\e_r =\; A^{-1}\er B^{-1}~,
      \\[0.1em]
      \ssr
      &\quad\longrightarrow\quad
      \tissr
      = C^{-1}\ssr B^{-1}~.
    \end{aligned}
  \end{cases}
  \label{eq:pmlcoefficients}
\end{align}
In the above, we introduced the $3\times3$ matrices
\begin{gather}
  A=T^{-1}_{\vec e_x\vec e_r}\text{diag}\,
  \Big(\frac{1}{\bar d^2},\frac{1}{d\bar d},\frac{1}{d\bar d}\Big)
  T_{\vec e_x\vec e_r}~,
  \quad
  B=T^{-1}_{\vec e_x\vec e_r}\text{diag}\,\big(d,\bar d,\bar d\big)
  T_{\vec e_x\vec e_r}~,
  \\\notag
  C=T^{-1}_{\vec e_x\vec e_r}\text{diag}\,\Big(\frac{1}{\bar d},\frac{1}{\bar d},
  \frac{1}{d}\Big)
  T_{\vec e_x\vec e_r}~,
\end{gather}
where
\begin{align}
  \label{eq:d}
  d=1+i\,s(r)~, \quad
  \bar d=1+i/r\int_\rho^r s(\tau)\,\text{d}\tau~,
\end{align}
for an appropriately chosen scaling factor $s(\tau)$ that will be defined
later. Note that $T_{\vec e_x\vec e_r}$ is the matrix that rotates $\vec
e_r$ onto $\vec e_x$, and $\tau$ is the distance from the origin. The PML
is assumed to be a spherical shell starting at distance $\rho$ from the
origin, as shown in Figure~\ref{fig:geometries}.

\subsection{Adaptive local refinement}
\label{subsec:adapt-local}

By the assumption that $\text{Im}\,\ssr>0$ in the nonretarded frequency
regime (Definition~\ref{def:nonr}), the SPP has a wavelength much smaller
than the one in the ambient medium (at the same frequency, $\omega$). Thus,
wave propagation along the metamaterial sheet, $\Sigma$, has a pronounced
two-scale character, being characterized by length scales of the order of
$1/(\text{Re}\,\km)$ and $1/k$; here, $\text{Re}\,\km\gg k$.

In our numerical simulation, we use typical values of $\ssr$ in the
nonretarded regime for which the SPP wavelength is one to two orders of
magnitude smaller than the wavelength in the surrounding medium. This poses
a challenge because, on the one hand, the minimal computational domain
(that still has a well-controlled error in slow oscillating modes) is
limited by the free-space wavelength, $2\pi/k$; on the other hand, the
minimal resolution necessary to resolve SPPs scales with $1/|\km|$.
Accordingly, in order to minimize computational cost while ensuring that
the SPP is sufficiently resolved, we use an adaptive, local refinement
strategy based on the \emph{dual weighted residual
method}~\cite{becker2001}.

Next, we outline the basics of our strategy. Starting from a relatively
coarse mesh, the resolution is successively improved by a number of
iterative refinement steps where a subset of cells is chosen for
refinement. The selection of cells for refinement is made with the help of
local (per cell $K$) error indicators, $\eta_k$, that are given by
\begin{align*}
  \eta_K^2\;=\;\rho_K^2\,\omega_K^2 + \rho_{\partial_K}^2\,\omega_{\partial
  K}^2~.
\end{align*}
The cell-wise residuals, viz., the integrals~\cite{becker2001,MML-jcp}
\begin{align*}
  \rho_K^2&=
  \int_K\big\|\nabla\times(\muri\nabla\times\vE_h)+
  \er\vE_h \big\|^2\,\text{d}x~,
  \\
  \rho_{\partial K}^2&=\frac12
  \int_K\big\|\big[\vn\times(\muri\nabla\times\vE_h)
  -i\ssr\vE_h\chi_\Sigma
  -i\sqrt{\muri\er}(\vE_h)_T\chi_\Omega
  \big]\big\|^2\,\text{d}o_x~,
\end{align*}
are multiplied by the \emph{weights}
\begin{align*}
  \omega_K^2 = \int_K\big\|z-I_hz\big\|^2\,\text{d}x,
  \quad
  \omega_{\partial K}^2 = \int_K\big\|z-I_hz\big\|^2\,\text{d}o_x~.
\end{align*}
In the above, $\vE_h$ denotes the finite element approximation on $\vE$.
The weights, $\omega_K$ and $\omega_{\partial K}$, are in turn computed
with the help of: the solutions, $z_K$, of a ``dual'' problem and their
respective interpolants, $I_hz$, in the finite element space. The rationale
of using a dual problem for computing the requisite weights is that these
can be tuned to a \emph{quantity of interest} in the form of a (possibly
non-linear) functional~\cite{becker2001}. In our particular application, we
choose to use
\begin{align}
  \mathcal{J}(\vE):=\frac12\int_\Sigma\big\|\nabla\times\vE\,\big\|^2
  \rm{d}o_x
  \label{eq:quantityofinterest}
\end{align}
as the quantity of interest. Hence, the dual problem reads
\begin{multline}
  \int_\Omega (\muri\nabla\times\vp)\cdot(\nabla\times\bar\vz)\dx
  \;\;-\;
  \int_\Omega(\er\vp)\cdot\bar\vz\dx
  -\,i\,\int_\Sigma(\ssr\vp_T)\cdot\bar\vz_T\dox
  \\
  \;-\;
  i\,\int_{\partial\Omega}\sqrt{\muri\er}\vp_T\cdot\bar\vz_T\dox
  \;=\;
  \int_\Sigma (\nabla\times\vE_h)\cdot(\nabla\times\bar\vp)\dx~.
  \label{eq:dualproblem}
\end{multline}

\section{Validation of numerical method and further numerics}
\label{sec:computation}
In this section, we focus on numerical computations by our finite element
method. These computations have a two-fold purpose: validation of our
numerical method and extraction of further predictions. First, we address
the prototypical geometry of the semi-infinite metamaterial sheet in order
to validate and verify our numerical approach by comparison of simulations
against the analytical description of section~\ref{subsec:integral_eqn}.
Second, we numerically simulate wave propagation under an incident plane
wave in a number of realistic geometries which are relevant to
nanophotonics applications; to our knowledge, no analytical results are
available for these geometries in the existing literature. For the
numerical experiments, we use $E_x^{\text in}= i$ (i.e., $E_{0}=i$ and
$\alpha=0$) throughout.

Figure~\ref{fig:geometries} depicts the geometries used in our numerical
tests. We examine the following configurations.
\begin{itemize}
  \item
    The reference case, i.e., the semi-infinite metamaterial sheet
    [Figure~\ref{fig:geometries}(a)]. Our numerical simulations for this
    setting are compared against an analytical solution. In particular, we
    verify the quality of approximation \eqref{eq:far-field-E} for the
    $x$-component of the electric field.
  \item
    Two co-planar, symmetrically placed, semi-infinite metamaterial sheets
    with a gap [Figure~\ref{fig:geometries}(b, c)]. In this configuration,
    the incident plane wave excites an SPP on each sheet. If the edges of
    the sheets are sufficiently close to each other, the induced SPPs may
    interfere destructively. We numerically examine the relative amplitude
    of the resulting SPP (compared to the reference case) on one sheet as a
    function of the gap width, $d$. We consider two different cases for the
    material of the gap. First, the gap is empty
    (Figure~\ref{fig:geometries}b); and, second, the gap is filled with a
    highly conducting material (Figure~\ref{fig:geometries}c). Accordingly,
    we show numerically that the dependence of the SPP amplitude on $d$ is
    dramatically different in these cases.
  \item
    A finite strip of metamaterial [Figure~\ref{fig:geometries}(d)]. For
    our numerics, the width, $d$, of the strip is chosen to have small to
    intermediate values compared to $1/k$. Because of a standing wave
    formed on the strip, we expect that a strong resonance effect can occur
    for suitable values of $d$. We numerically verify this resonance and
    quantify the (maximal) SPP magnitude as a function of $d$.
\end{itemize}

\subsection{Semi-infinite strip}
\label{subsec:semi-inf-num}

\begin{figure}[!tbp]
  \centering
  \subfloat[$\text{Re}\,E^{\text{sc}}_x$]{
    \includegraphics{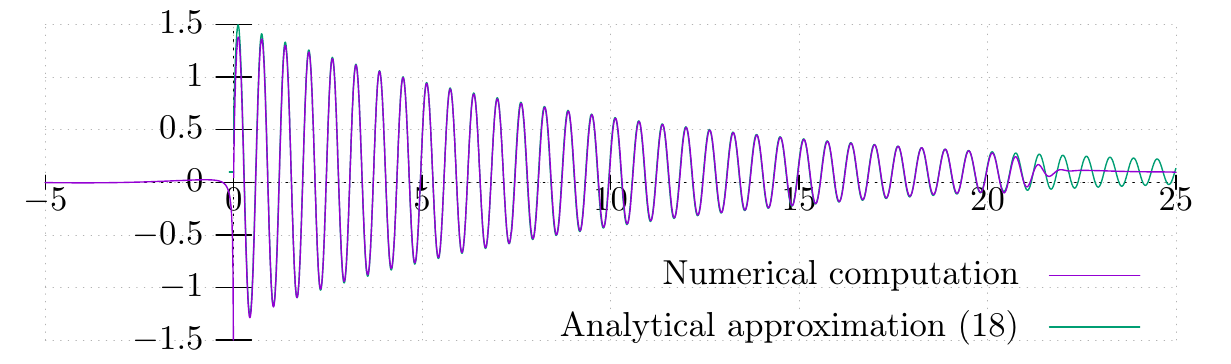}
  }

  \subfloat[$\text{Im}\,E^{\text{sc}}_x$]{
    \includegraphics{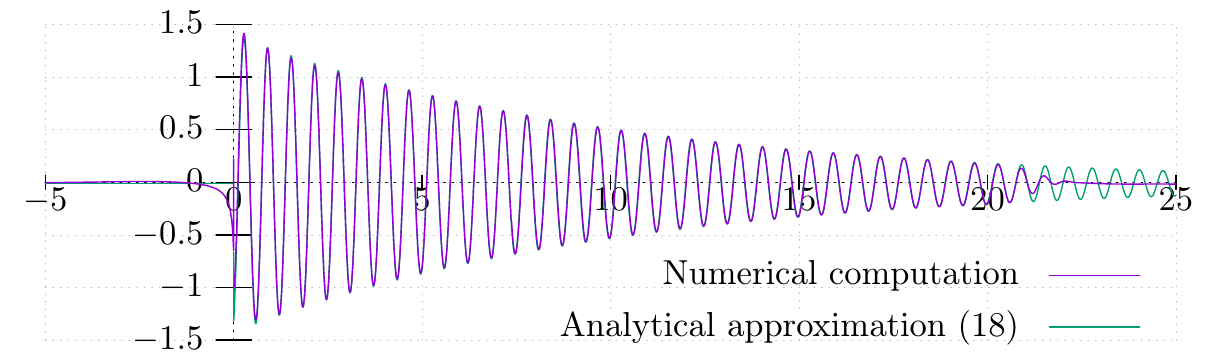}
  }

  \subfloat[$\text{Re}\,E^{\text{sc}}_x$]{
    \includegraphics{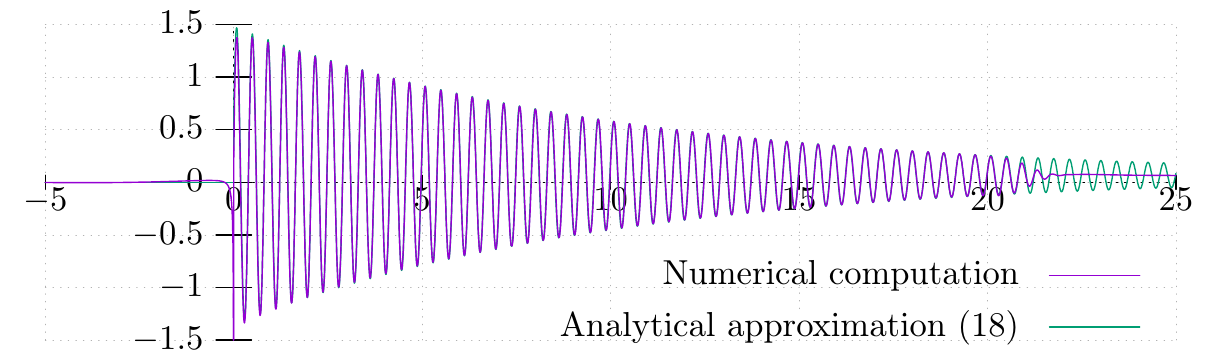}
  }

  \subfloat[$\text{Im}\,E^{\text{sc}}_x$]{
    \includegraphics{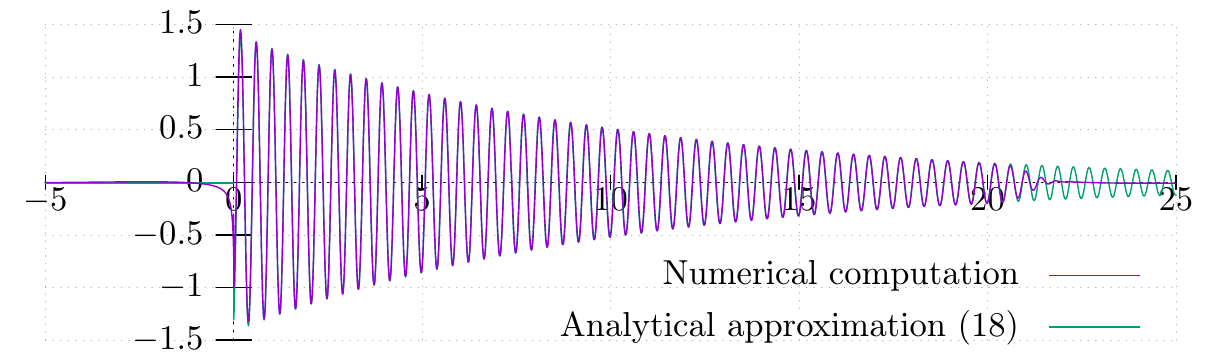}
  }
  \caption{[Color online]
    Real and imaginary parts of the $x$-component of scattered
    electric field, $E_x^{\rm sc}$, versus spatial coordinate, $x$, on
    semi-infinite sheet, $\Sigma=\{(x,0)\in\mathbb{R}^2\,:\,x>0\}$.
    The plots depict results based on: our numerical method;
    and analytical formula~(\ref{eq:far-field-E}), or~\eqref{eq:far-field-E2}.
    The values of the
    (rescaled) surface conductivity are: $\ssr=2.0\cdot10^{-3}+0.2i$ (a, b) and
    $\ssr=8.89\cdot10^{-4}+0.133i$ (c, d).}
  \label{fig:semi-infinite}
\end{figure}
\begin{figure}[!tbp]
  \centering
  \includegraphics{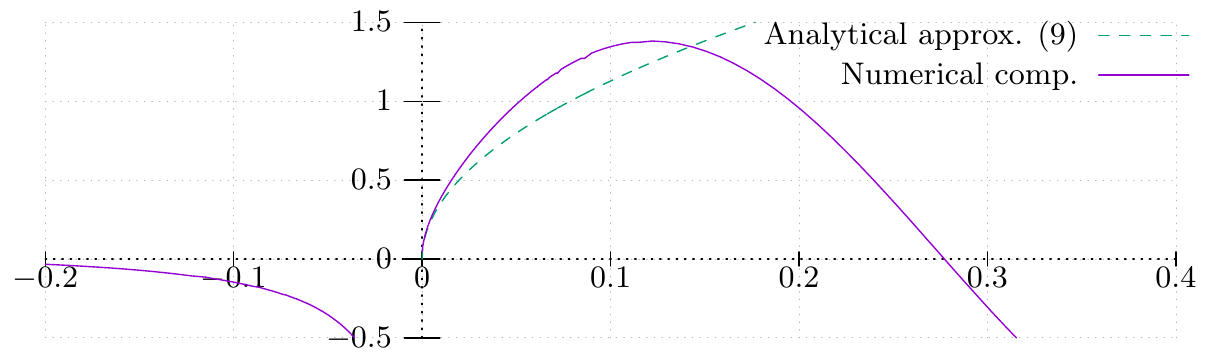}
  \caption{[Color online]
    Real part of the $x$-component of (total) electric field, $E_x$, near
    the edge of the sheet versus spatial coordinate, $x$. The value of
    (rescaled) surface conductivity is $\ssr=2.0\cdot10^{-3}+0.2i$. Our
    numerical computation (solid line) is compared  against near-edge
    asymptotic formula \eqref{eq:asymptotic0} (dashes). In the limit as
    $x\to 0$, the numerical and analytical results are in good agreement.
  }
  \label{fig:near-field-behavior}
\end{figure}
First, we consider the geometry of the semi-infinite strip,
$\Sigma=\{(x,0)\in\mathbb{R}^2\,:\,x\ge 0\}$, depicted in
Figure~\ref{fig:geometries}(a). The primary purpose of our numerical
simulation in this setting is to validate the analytical results discussed
in section~\ref{subsec:integral_eqn}. In particular, by subtraction of the
incident field, approximate formula~\eqref{eq:far-field-E} yields a
corresponding expression for the $x$-component of the scattered electric
field, $E^{\text{sc}}_x$. In terms of the rescaled quantities
(section~\ref{subsec:var-form}), the scattered field reads
\begin{align}
  E^{\text{sc}}_{x,r}(x) \;\simeq\;
  E_0\,\left[\frac{1}{1+\frac{\ssr}{2}}-1\right]
  -E_0\,e^{-\hQ_+(\sin\alpha)+\hQ_+(\kmr)}e^{i\kmr x}~,
  \label{eq:far-field-E2}
\end{align}
where the first term (with brackets) represents the directly reflected
field and the second term is the SPP. Here, for $\alpha=0$, the requisite
values of $\hQ_+(\xi)$ in the exponent are $\hQ_+(0)\approx0.0005+0.05i$,
and $\hQ_+(k_m/k)\approx0.346574+0.392699i$. Note that in the case with the
surface conductivity $\ssr=2.0\cdot10^{-3}+0.2i$
($\ssr=8.89\cdot10^{-4}+0.133i$), the SPP wavenumber is
$\kmr\;=\;10.0489+0.0994937i$ ($\kmr\;=\;15.0701+0.100288i$).

In Figure~\ref{fig:semi-infinite}, we compare graphically the outcome of
formula~\eqref{eq:far-field-E2} to results of our numerical method for the
corresponding scattered field. Specifically, the real and imaginary parts
of the scattered electric field in the $x$-direction, $E_x^{\rm sc}$, are
plotted as a function of the spatial coordinate, $x$, along the sheet,
$\Sigma$. Evidently, the numerical and analytical results are in excellent
agreement outside the PML and for (roughly) 2-3 SPP wavelengths away from
the origin ($2\le x\le 20$ in our numerics).

Furthermore, we test the finite element numerical simulations against
analytical prediction~\eqref{eq:asymptotic0} for the asymptotic behavior of
$E_x$ along the sheet near the edge, as $x$ becomes sufficiently small. The
predicted behavior is confirmed numerically in
Figure~\ref{fig:near-field-behavior}, in which the real part of this field
component is plotted versus $x$ closely enough to the edge of the sheet.

\subsection{Co-planar, semi-infinite sheets with a gap}
\label{subsec:two-sheets-gap}

\begin{figure}[!t]
  \centering
  \includegraphics{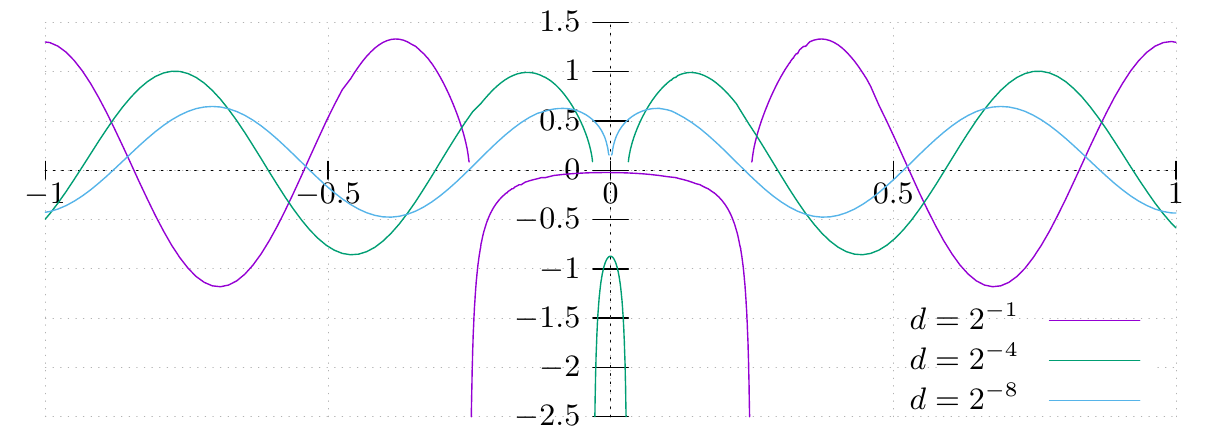}
  \caption{[Color online]
      Real part of the $x$-component of the scattered electric field,
      $E_x^{\rm sc}$, as a function of the spatial coordinate, $x$, in the
      case with two co-planar, semi-infinite sheets having an empty gap in
      the interval $[-d/2, d/2]$. The plots show numerical results for
      three different values of the gap width: $d=2^{-1}$, $2^{-4}$, and
      $2^{-8}$.}
  \label{fig:gap-zoom}
\end{figure}
\begin{figure}[!t]
  \centering
  \includegraphics{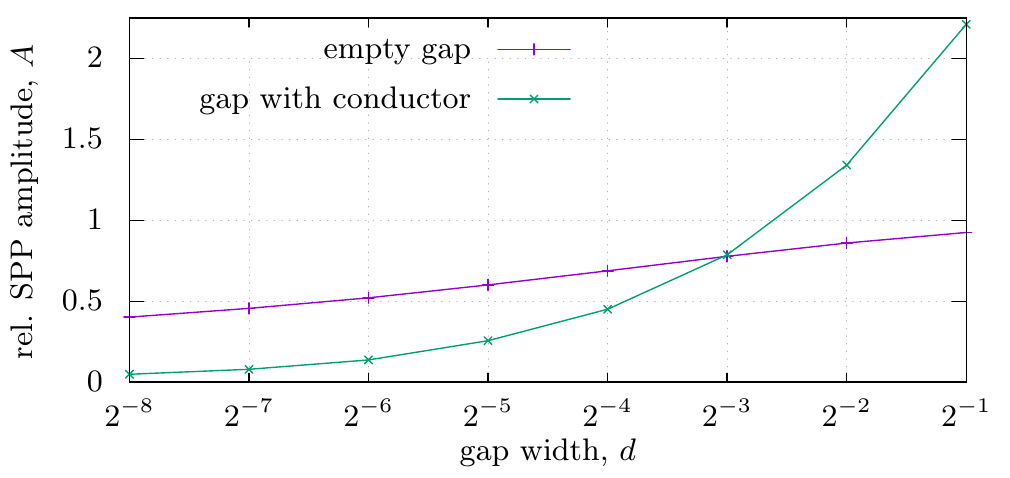}
  \caption{[Color online]
    Relative amplitude, $A$ (compared to the reference case) of the SPP
    versus gap width, $d$, according to formula~\eqref{eq:SPP-paramet} in
    the geometry with two co-planar, semi-infinite sheets. The amplitude
    $A$ has been computed via fitting of formula~\eqref{eq:SPP-paramet} to
    numerical solution for scattered field.}
  \label{fig:gap-functionaldep}
\end{figure}
Next, we examine the (symmetric) configuration that consists of two
co-planar, semi-infinite sheets with a gap of width $d$,
$\Sigma=\{(x,0)\in\mathbb{R}^2\,:\,x<-d/2\ \mbox{or}\ x>d/2\}$; see
Figure~\ref{fig:geometries}(b,c). This setting is of fundamental interest
in applications because experimental setups often involve arrays of strips
of metamaterials such as graphene~\cite{bludov13}. By using the present
geometry, we are able to isolate the influence of the gap width on the
surface plasmon of the metamaterial, since each sheet is infinitely long.
In particular, resonances associated with standing waves on the sheet are
absent from this setting (see section~\ref{subsec:strip-num} for the
resonant, finite strip).

We consider two cases for the material of the gap. In one case, the gap is
empty, or, in other words, the material of the gap is identified with the
medium of the ambient space [Figure~\ref{fig:geometries}(b)]; and in
another case the gap is filled with a highly conducting material
[Figure~\ref{fig:geometries}(c)]. Because of the proximity of the two
edges, the SPPs induced on each sheet are in principle expected to
interfere. Our numerics show that the closer the sheets are to each other,
i.e., as the gap width, $d$, decreases, the stronger the effect of
destructive interference of the SPPs is. For example, this trend is
demonstrated in Figure~\ref{fig:gap-zoom}, which depicts the SPP versus $x$
near the empty gap for three values of $d$.

We perform a parameter study in order to examine the strength of the SPP as
a function of the gap width. We examine the cases with an empty gap and a
gap filled with a material of (scaled) surface conductivity
$\sigma_{r,2}^\Sigma$ such that $\text{Re}\,\sigma_{r,2}^\Sigma \gg 1$,
$\text{Im}\,\sigma_{r,2}^\Sigma=0$. In this study, we carry out the fitting
of the finite element-based numerical solution for the $x$-component of the
scattered electric field to an analytical formula constructed
from~\eqref{eq:far-field-E2}. Specifically, this formula reads
\begin{align}\label{eq:SPP-paramet}
  E_{x,r}^{\rm sc}(x)\approx E_0\, \left[\frac{1}{1+\frac{\ssr}{2}}-1\right]
  -A\,E_0\,e^{-\hQ_+(\sin\alpha)+\hQ_+(\kmr)}e^{i(\kmr x+\varphi)}~,\quad x>0~,
\end{align}
which is supplemented with two parameters, namely, the extra factor, or
{\em relative amplitude}, $A$ for the SPP and a corresponding phase shift,
$\varphi$. These parameters, $A$ and $\varphi$, are determined via fitting
of the last expression for $E_{x,r}^{\rm sc}$ to our numerics.

Interestingly, in the two distinct cases of the gap material mentioned
above, we observe two fundamentally different scaling laws for $A$, which
hold for a range of values of $d$. These behaviors are shown in a
semi-logarithmic plot of the relative amplitude, $A$, versus the gap width,
$d$; see~Figure~\ref{fig:gap-functionaldep}. In the case with an empty gap,
we observe that the relative amplitude, $A$, exhibits a logarithmic
behavior with $d$, viz., $A \sim -\log\,d$. In contrast, in the case with a
highly conducting material, the corresponding scaling is approximately
linear, viz., $A \sim d$. Both scaling laws hold at least for $2^{-8}\le
d\le 2^{-4}$ (Figure~\ref{fig:gap-functionaldep}). Outside this region for
$d$, our numerical fitting yields results consistent with the limits
$\lim_{d\to\infty} A = A_{\infty}$ and $\lim_{d\to 0} A= 0$, where
$A_\infty$ is the (constant) relative amplitude of the SPP for an
infinitely large gap, when the two sheets tend to be isolated from each
other.

\subsection{Finite strip}
\label{subsec:strip-num}

\begin{figure}[!t]
  \centering
  \subfloat[]{
    \includegraphics{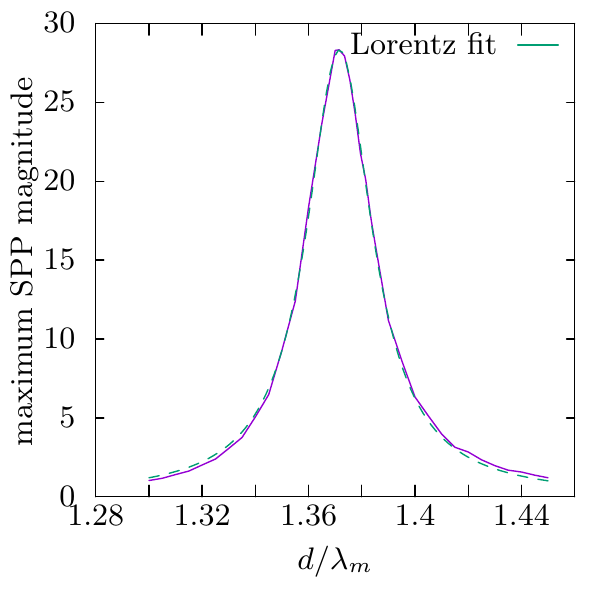}
  }
  \subfloat[]{
    \includegraphics{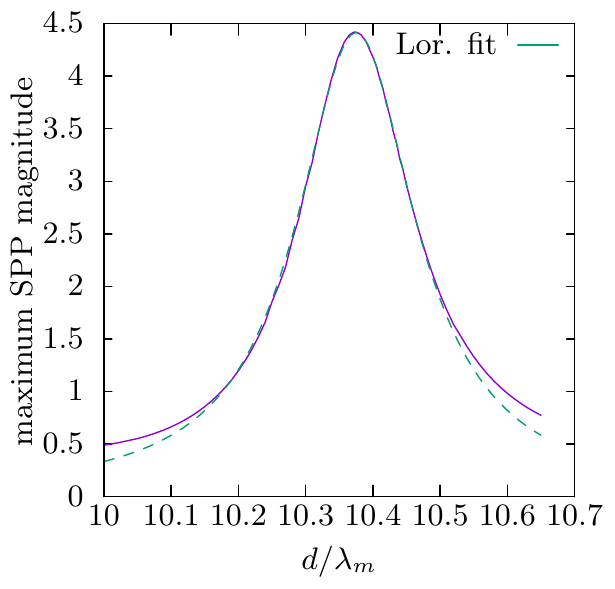}
  }
  \caption{[Color online]
    Maximal magnitude of SPP as a function of strip width, $d/\lambda_m$,
    relative to SPP wavelength, $\lambda_m=2\pi/(\text{Re}\,\km)$. Two
    distinct maxima are shown, corresponding to the first [(a)] and tenth
    [(b)] resonance observed in the chosen range of $d$. The numerical results
    (dashes) are fitted to a Lorentzian function (solid line).}
  \label{fig:resonance}
\end{figure}
Next, we consider the geometry of a finite metamaterial strip,
$\Sigma=\{(x,0)\in\mathbb{R}^2\,:\,-d/2\le x\le d/2\}$
[Figure~\ref{fig:geometries}(d)]. In this case, if the length $d$ of the
strip is sufficiently large, the two edges are expected to cause resonances
to the observed SPP. In other words, the maximal magnitude of the SPP
generated in this structure should exhibit peaks as a function of the strip
width, $d$ (for fixed $k$).

In our numerical simulations, we compute the resulting, maximal magnitude
of the SPP as a function of $d$; see~Figure~\ref{fig:resonance}. For these
numerics, we choose the surface conductivity to be equal to
$\ssr=2.0\cdot10^{-3}+0.2i$. We select two peaks among those observed in
the chosen range of values of $d$. These resonances are depicted in
separate plots here [Figure~\ref{fig:resonance}(a,b)]. In order to quantify
the two selected resonances, we carry out the fitting of the numerical
results in each case to the Lorentzian function
\begin{align*}
  \varphi(d) =
  \frac{A\,\pi}{\gamma}\,\bigl[1+(d-d_0)^2/\gamma^2\bigr]^{-1}~.
\end{align*}
The parameters $d_0$, $\gamma$ and $A$ of this formula are  determined via
the fitting procedure. Specifically, we obtain the following values:
$d_0=1.372\,\lambda_m$, $\gamma=0.015\,\lambda_m$, $A=1.34$ for the first
resonance [Figure~\ref{fig:resonance}(a)]; and $d_0=10.375\,\lambda_m$,
$\gamma=0.11\,\lambda_m$, $A=1.49$ for another (tenth) resonance
[Figure~\ref{fig:resonance}(b)] in the chosen range of values of $d$. Here,
$\lambda_m=2\pi/(\text{Re}\,k_m)$ is the SPP wavelength on the
semi-infinite sheet of the same conductivity (see
section~\ref{subsec:integral_eqn}).

\section{Conclusion}
\label{sec:conclusion}

In this paper, we numerically studied the generation of SPPs on atomically
thick metamaterial sheets by edge effects in different geometries. Our
chosen configurations included: a semi-infinite sheet; two co-planar,
semi-infinite sheets with a gap of variable width; and a finite strip of
variable width. In our computations, we used an adaptive finite element
method with curl-conforming Nédélec-elements in order to resolve the fine
scale of the SPP propagating along the sheet in the presence of edges. Our
numerical approach here forms an extension of the method introduced
in~\cite{MML-jcp} to more realistic geometries.

We validated our numerical treatment by comparison of the finite
element-based numerics to an analytical solution for the semi-infinite
sheet~\cite{MML-sapm}. By further numerical simulations, we demonstrated
that the presence of a highly conducting material in the gap between two
co-planar, semi-infinite metamaterial sheets can increase the SPP
magnitude, and leads to a distinctly different dependence of the SPP on the
gap width in comparison to the case with an empty gap. In addition, we
numerically characterized SPP resonances on a finite strip, and
demonstrated that the SPP maximal magnitude (as a function of gap width) is
well described by a Lorentzian function.

Our results point to a few open problems. For instance, our computations
have focused on 2D, although our approach is applicable to arbitrary
spatial dimensions. An emerging question concerns the use of truly
three-dimensional (3D) geometries, which may contain metamaterial sheets
with corners, conical singularities or arbitrarily curved hypersurfaces.
The generation and propagation of SPPs in 3D settings is the subject of
ongoing work.


\bibliographystyle{siamplain}

\end{document}